**Development of a Novel Computational Model for**

**Evaluating Fall Risk in Patient Room Design**

Roya Sabbagh Novin[1], Ellen Taylor[2], Tucker Hermans[3], Andrew Merryweather[1]

[1] Department of Mechanical Engineering and Robotics Center, University of Utah, USA
[2] The Center for Health Design, Concord, CA, USA
[3] School of Computing and Robotics Center, University of Utah, USA

**Author Note**

Roya Sabbagh Novin:   https://orcid.org/0000-0003-4265-0661

Correspondence concerning this article should be addressed to Roya Sabbagh Novin, Department of Mechanical Engineering and Robotics Center, University of Utah, Utah, USA. roya.sabbaghnovin@utah.edu

The authors disclosed receipt of the following financial support for the research, authorship, and/or publication of this article: This work was supported by AHRQ under award number [1R18HS025606].

The authors declare that there is no conflict of interest

## Abstract

**Objectives:** This study proposes a computational model to evaluate patient room design layout and features that contribute to patient stability and mitigate the risk of fall.

**Background:** While common fall risk assessment tools in nursing have an acceptable level of sensitivity and specificity, they focus on intrinsic factors and medications, making risk assessment limited in terms of how the physical environment contributes to fall risk.

**Methods:** We use literature to inform a computational model (algorithm) to define the relationship between these factors and the risk of fall. We use a trajectory optimization approach for patient motion prediction.

**Results:** Based on available data, the algorithm includes static factors of lighting, flooring, supportive objects, and bathroom doors, and dynamic factors of patient movement. This preliminary model was tested using four room designs as examples of typical room configurations. Results show the capabilities of the proposed model to identify the risk associated with different room layouts and features.

**Conclusions:** This innovative approach to room design evaluation and resulting estimation of patient fall risk show promise as proactive evidence-based tool to evaluate the relationship of potential fall risk and room design. The development of the model highlights the challenge of heterogeneity in factors and reporting found in the studies of patient falls, which hinder our understanding of the role of the built environment in mitigating risk. A more comprehensive investigation comparing the model with actual patient falls data is needed to further refine model development.

**Executive Summary of Key Concepts:**

The role of the environment is recognized as a factor that contributes to the risk of falls in hospital patient rooms, but there is not adequate empirical data to support evidence-based decision-making for stakeholders. We propose a computational model for fall risk evaluation of patient rooms. The proposed model consists of two major parts: (1) the room baseline evaluation, which considers room design and layout as static factors; and (2) motion evaluation, which includes dynamic factors based on expected patient movements within the room.

We focus on factors defined in the physical environment and believe that in conjunction with other fall risk assessment tools, we will gain greater knowledge of the risk of falls in a hospital patient room.

The comprehensive fall risk model and its application is a significant step in understanding and solving the problem of patient falls in hospitals. It can be used to provide guidance for healthcare decision makers to optimize effective environmental interventions to reduce risk of falls while promoting safe patient mobility in the hospital room environment. The model is structured to allow for integration with other competing demands in a patient room to achieve an optimal room design.

**Implications for Practice:**

- Results from published research can often be difficult for stakeholders to interpret and incorporate into actionable decision-making, especially when the data are incomplete, heterogeneous, or inconsistent and risk is not well understood. Outlining defined risk assumptions for specific design interventions can advance our understanding of the role of the environment in patient safety, specifically patient falls.

- The development of computational models can enable stakeholders to simultaneously evaluate the risk of individual design decisions along with the probable activities of the patient, advancing a structured, proactive process to evaluate risk during design.

- Visual heat maps, generated as a result of evidence-based computational models, can allow stakeholders to more easily understand the implications of specific design decisions.

- The preliminary development of a computational model suggests that intuitive assumptions about design decisions need to be balanced by the details in those decisions relative to patient activity, including walking distances (irrespective of headwall or footwall bathroom location), supportive features (e.g., grab bars) for sit-to-stand, stand-to-sit and gait activities, and locations of furniture and fixtures that minimize turning or other suboptimal biomechanics that pose a risk for patient falls. The flexibility of the proposed computational model will allow the tool to be updated as evidence becomes available, with factors that can be fine-tuned for project-specific design objectives.

## Development of a Novel Computational Model for

## Evaluating Fall Risk in Patient Room Design

### Introduction

Patient falls are a continued focus for improving the quality of healthcare (AHRQ, 2016). Falls are nationally reported in the US as a hospital-acquired condition (HAC). Falls with serious injury were recorded as the most reported sentinel event in 2018 (The Joint Commission, 2019) and on an average, cost about $14,000 (Haines et al., 2013). In addition to unnecessary financial costs to both patients and the healthcare system, falls increase morbidity, length of stay, and reduce quality of life. The tension of preventing falls, while promoting mobility, mandates identification of both static and dynamic factors that contribute to patient stability and reduce risk of falls.

The major risk factors for patient falls are categorized as (1) intrinsic factors such as age, falls history, and physical impairment, (2) extrinsic factors such as room configuration, assistive equipment, and medications (Choi, 2011). Most of the literature on patient falls focus on the intrinsic factors or medications. Very few empirical studies investigate factors related to the physical environment (Anonymous, 2016), and according to our own review, none of the commonly used fall risk assessment tools include built environment factors (Pati, Lee, Mihandoust, Kazem-Zadeh, & Oh, 2018). The most commonly used fall risk assessment tools are the Hendrich II Fall Risk Model (Hendrich, 2013), the Morse Fall Scale (Morse, Morse, & Tylko, 1989), and the St. Thomas Risk Assessment Tool (Oliver, Britton, Seed, Martin, & Hopper, 1997). **It would be equally beneficial to have a detailed and systematic fall risk evaluation that includes environmental features that can be used for proactive evaluation of risk as part of the patient room design process.** In this paper, we propose a novel computational model for fall risk evaluation of patient room design (Figure 1.). **We focus on factors defined by the physical** **environment and believe that in conjunction with other fall risk assessment tools, we can gain greater knowledge of the risk of falls in a hospital patient room.**

[Place Figure 1 approximately here]

A critical step in developing a more comprehensive fall risk evaluation tool is to define the contributing built environment factors and estimate their effects on the fall. Each factor has unique properties and characteristics which should be taken into account. However, current empirical research on environment-related factors is limited. In this paper, we use the limited data available in the literature for each contributing factor, and make assumptions as outlined in the narrative description. In other words, the proposed model should be treated as a framework for future studies and a tool that can be used to simultaneously investigate the influence of a variety of contributing factors related to falls. The proposed model, which is the first phase of a larger five-year project, consists of: (1) the room baseline evaluation, considering *static* factors based on room layout; and (2) trajectory evaluation, considering *dynamic* factors based on the expected patient motion.

## Background

Fall risk assessments provide a structured evaluation of factors that may increase a patient's risk of falling (Aranda-Gallardo et al., 2013).  However, most fall risk assessments only consider intrinsic factors and medications, making the prediction limited in terms of how the physical environment contributes to fall risk. In addition, a large percentage of patients who fell were scored as low risk, showing these commonly used risk assessment tools fall short and that the specific population and setting affect the performance of these assessments (Swartzell et al., 2013).

In the following, we summarize the current state of literature on factors of the physical environment which are included in our model. The reason that we focus on these specific factors are: (1) there is a causality relationship shown between these factors and risk of fall; (2) there is literature showing the effect of these factors that can be used to inform the model; and (3) as more factors are added to the model, the validation process becomes more complicated, and a future stepwise approach would be more feasible and reliable for advancing model validity. Although the literature is not complete, even for these factors, we believe that it is enough to establish an initial model to be further validated by future human subject studies.

There are additional factors that are not included in the current model due to their complexity and lack of studies to establish the magnitude of their effects (e.g. patient bed visibility has been reported as a factor (Bosch et al., 2016; Choi, 2011)). For these factors, the causality relationship is not clear enough to conduct solid assumption. The flexible design of our fall risk model allows for the addition of more factors as data become available through future work.

**Flooring Factors**

Several studies on fall events evaluate specific flooring (Warren & Hanger, 2013; Calkins, Biddle, & Biesan, 2012; Mackey et al., 2019). In general, there is no statistically significant difference between number of falls for different floor types in the patient room. Even the literature showing a difference has limitations since other factors might have contributed to differences (Choi, 2011). Warren & Hanger (2013) found a statistically significant higher rate of falls on carpet as compared to vinyl in the psychiatric ward, but no significant difference in fall rates between the two materials in stroke and general wards. Donald et al. (2000) found that the number of falls on carpet is more than on vinyl floors, but the results were not statistically significant, and the time period of this study was relatively short (nine months). Transitioning between flooring surfaces with different coefficients of friction was shown to increase risk of trip induced fall, especially among older adults (Kim & Lockhart, 2019). On a different floor type characteristic, Calkins et al. (2012) show that having a medium-size pattern (1″- 6″) was associated with greater falls than no pattern, small pattern ($\leq 1''$) or large pattern ($\geq 6''$). However, due to inconsistency of the number of rooms for each pattern, it is not possible to make a concrete conclusion.

## Lighting Factors

Literature on the effect of lighting on falls is often non-descript, citing "poor lighting" as a contributing factor to fall. Several studies focus on gait analysis in various light intensities. To the best knowledge of the authors, there has been no empirical research studying the direct effect of lighting on risk of falls in hospital patient rooms. Figueiro et al. (2008) show in a study that participants reduced their velocity from 85.6 cm/s (1.91 mph) to 75.6 cm/s (1.69 mph), when ambient lights were turned off and night lights were turned on. Verghese et al. (2009) propose a 10 cm/sec (0.22 mph) drop in gait velocity is associated with a 7% higher risk of falls.

## Supporting Objects

Object properties are categorized as: (1) height of the supporting surface, (2) movability, and (3) graspability. Komisar et al. (2019) show that handrail height did not significantly affect handrail contact or movement time. This may have resulted from differences in reaching strategy, including increased deceleration times, allowing for greater reach control without consequence to contact time. Movability has two parts: (1) how easily the whole object moves when applying a force (e.g., bed rail movement or an object on wheels), and (2) the object's compliance (softness), such as seat cushion or mattress. To the best knowledge of the authors, there is no literature on the effect of object immovability on supporting levels for different objects, however experience suggest that an object that is easily movable provides poor support.

Graspability has two aspects: (1) how well an object can be grasped, and (2) the grasping surface friction. Dam-Huisman et al. (2015) studied the effect of handrail cross-section on lateral falling and comfort during support and found that smaller cylindrical handrails (diameter of $1.1''$ and 1.26″) are more suitable to apply high forces on than larger ones (1.77″), oval handrails or squared handrails. Young et al. (2012) studied the effect of handhold orientation, size, and wearing a glove on the maximum breakaway strength. They show that breakaway strength increased 75% to 94% as the handhold orientation was moved from vertical to horizontal. Breakaway strength decreased 8% to 13% for large diameter (2″) handholds compared to smaller ones (0.86″ to 1.26″). Gloves with greater friction increased breakaway force.

**Door Operation**

In a descriptive analysis of hospital falls, it is suggested that door widths (Tzeng, 2011), adjacent clearances (Calkins et al., 2012), and swing direction (Pati et al., 2017; Calkins et al., 2012) force postural changes which affects the risk of fall (Tzeng, 2011; Pati et al., 2018). They suggest there are more falls associated with bathroom doors that open into the bathroom than swing outwards. (Outward swinging door also allows staff to enter the room after a fall.)

**Activity Factors**

The majority of patient movement in a hospital setting requires a Sit-to-Stand and walking. It is well known that Sit-to-Stand puts greater strength, postural control and vestibular demands on a patient and has implications for fall risk (Fujimoto & Chou, 2012; Galán-Mercant & Cuesta-Vargas, 2014). Literature on the effect of type of activity on fall risk suggest that in a bathroom setting, motion-related factors turning, pushing, pulling, and grabbing are significant. For the clinician zone, motion-related factors of pushing and pulling demonstrated statistical significance.

**Methods**

To develop a comprehensive fall risk assessment model, we categorize the contributing factors into static and dynamic components. Figure 2 shows a framework of our proposed model. Based on the static factors of floor type, lighting, door operation and the effect of supporting objects, we find the room baseline evaluation which represents risk distribution over the entire room. These are shown in Figure 3 as heat maps on a grid map of the room for each individual factor and the baseline evaluation. These heat maps show the percentage of increase or decrease in the fall risk at each grid. Any grid with values less than 1.0 (blue grids) means the fall risk is decreased. In other words, some sort of support is provided in those areas. Values more than 1.0 (red grids) show that there is high fall risk and more support is needed. We provide an example of calculations for a certain grid for more clarification.

[Place Figure 2 approximately here]

Dynamic factors include patient motion such as gait properties and type of activity. we define a set of common scenarios such as transition from bed to toilet and predict trajectories for all scenarios based on two main assumptions: (1) human motion is optimal and (2) the frail-elderly (or patients with pain or severe health conditions) move closer to external supporting points. For each point on the trajectory, we find the fall risk using the calculated baseline and the effect of dynamic factors. Finally, we combine the baseline and motion-based evaluation to obtain the overall fall risk for the entire room. For each factor (e.g. lighting, flooring), we define a function to represent the effect of that factor as a percentage of increase or decrease in fall risk. The parameters (coefficients) used to express these functions are assigned based on the literature.

[Place Figure 3 approximately here]

**Room Baseline Evaluation**

For visualization, we represent the entire room as a grid map. The size of the grid cells is somewhat arbitrary, we use 0.1m×0.1m (approximately 4″×4″) cells for this paper. For each cell, we find the effect of each factor on fall risk based on the associated models. The product of the effects of all static factors determines the level of baseline fall risk:

$$S_b(\chi) = \prod_{i=1}^{n} f_{\theta_i}(x_i) \qquad \textit{(Eq. 1)}$$

Here, n is the number of factors used in the model and $f_{\theta_i}(x_i)$ is a function of factor $x_i$ with function parameters $\theta_i$. These functions can be linear or nonlinear, continuous or discrete, and are defined based on the nature of the effect of that feature on fall risk. For example, since having an external support point that is not close enough to the patient has no effect on the patient's balance, we use a truncated linear function over distance to present the effect of that external support point. Functions for each factor involved in the baseline evaluation are described below.

### *Flooring*

We consider two types of flooring effect. First, for each grid we have a specific floor type with an associated fall risk score. Then we check for floor type transition in four major directions, if there is one, the fall risk is increased based on the following function:

$$\boldsymbol{f_{floor} = 1 + (c_i + \sum_{j=1}^{m} n_j c_{ij})} \qquad \textit{(Eq. 2)}$$

Where, $c_i$ is a constant value showing the increase/decrease in the fall risk based on the floor surface. We add the effect of transition by summing the effects of moving in four major directions. $m$ is the number of different floor types in the room, we count the number of directions that makes the transition to that type, $n_j$, and multiply it by the effect of that transition, $c_{ij}$. In this phase of tool development, two different types of floor surface are chosen to show how it would affect the risk of fall (Mackey et al., 2019).

[Place Table 1 approximately here]

### *Lighting*

We define three levels of light intensity: (1) High or ambient light ($\geq$ 500lux), which has no effect on the fall risk. (2) Low or night light ($\leq$ 100lux), which increases the fall risk by 7%. (3)

Medium light (anything between ambient and night light), which increases the risk of fall by 3% (Figueiro et al., 2008; Verghese et al., 2009). Therefore, the function is defined as a case function:

$$f_{light} = \begin{cases} 1.07, & x < l_1 \\ 1.03, & l_1 \leq x \leq l_2 \\ 1, & x \geq l_2 \end{cases} \qquad \textit{(Eq. 3)}$$

We find the intensity of light in lux for each grid based on the distance to all light sources. Then, based on the range it falls into, we assign the percentage of increase in fall risk. In order to show the effect of lighting, we compare fall risk evaluations for both day and night light versions.

***Supporting Objects***

There are two aspects in supporting level: the distance to the closest supporting object and the level of support of that object. For the distance to a supporting point we consider three ranges of distance based on type of reaching (Fromuth & Parkinson, 2008): (1) close, (2) reachable, (3) not reachable. The effect of a supporting object is maximum when it is close enough to the patient to grasp it with minimal postural change. As the distance increases, the patient has to modify posture for a successful grasp, decreasing the supporting effect of that object until it is out of reach and has no supporting effect at all. This is summarized in the following truncated linear function and shown in Figure 4 (coefficients are presented in Table 1).

$$O^* = \arg\min_{O} \|x - O\|^2 \qquad \textit{(Eq. 4)}$$

$$d = \|x - O^*\|^2 \qquad \textit{(Eq. 5)}$$

$$f_{support} = \begin{cases} \dfrac{\theta_1}{S_{O^*}}, & d < d_1 \\ \dfrac{(\theta_1 + \dfrac{(\theta_2 - \theta_1)}{(d_2 - d_1)}(d - d_1))}{S_{O^*}}, & d_1 \leq d \leq d_2 \\ \dfrac{\theta_2}{S_{O^*}}, & d \geq d_2 \end{cases} \qquad \textit{(Eq. 6)}$$

We first find the closest object $O^*$ and its distance to the grid. Based on the distance we choose the case for the support function. The numerator defines the effect of distance while the denominator incorporates the support level of the object. As the support level increases from one, the risk of fall decreases. The close distance $d_1, is$ assigned based on average human arm length. The reachable distance which defines $d_2$ is assigned based on maximum reach distance literature for a 50[th] percentile female (Tilley, 2001; Tantisuwat, Anong, Dannaovarat Chamonchant, 2014).

The more challenging part is assigning the level of support for the object. We have defined three main characteristics of an object that indicate the level of support: (1) height of the grasping point, (2) movability, and (3) graspability of the object. Based on these characteristics, we assign a support score to each object. Based on a preliminary survey from healthcare providers who have regular physical interaction with patients, all factors contribute equally. However, a detailed subject study is being conducted to define a more accurate function based on these three characteristics. Objects can also have negative effect on support level (<1.0). For example, objects with high movability such as over bed tables and IV poles, which are usually on wheels, do not provide stable support, but patients might lean on them potentially increasing the risk of a fall. We have assigned support levels between 0.6 (for a movable IV pole) to 1.3 (for a fixed hand rail).

[Place Figure 4 approximately here]

### Door Operation

We model doors with the swing direction inward or outward from the bathroom and also sliding doors. We estimate the coefficient based on the literature (Tzeng, 2011; Pati et al., 2018). In addition, there are differences between door widths - larger openings (>36") and smaller doors ($\leq 36$"). Based on inferences from correlational study (Calkins et al., 2012), we assume the wider door is slightly less risky than the narrow one, as it allows space for moving assistive devices like IV poles. The amount of increase in the fall risk based on the door types are provided in Table 1.

**Motion Evaluation**

To include patient motion effects, we define a set of scenarios that are common in a hospital patient room and assign a number of trajectories per scenario to account for task frequency (Tzeng & Yin, 2012). In order to have evaluation over the entire room, although the main scenario in a hospital room is bed to toilet or toilet to bed, we also include other scenarios, such as patient chair to the toilet, but with a lower frequency. Scenarios used in this research are provided in Table 2 along with the number of sample trajectories generated for each of them.

[Place Table 2 approximately here]

***Patient Motion Prediction***

To generate each trajectory, we sample two points, a point close to the initial location as the start state and a point close to the target location as the end state (Figure 5 (a)). We use a common optimization-based method to find a trajectory based on the current state and the desired state (Figure 5 (b)) (Ratliff, Zucker, Bagnell, & Srinivasa, 2009; Anonymous, 2018). We assume that patients will take a minimum length path and avoid obstacles during walking. We additionally believe that patients will try to avoid falling by moving closer to external support points when available. Assuming patients perform optimal motions with respect to the path length and risk of fall assumption, we can formulate a patient trajectory prediction as an optimization problem:

$$\min_{\zeta^p} \quad J = \sum_{t=1}^{h} \left( \lambda \left\| \zeta_t^p - \zeta_g^p \right\|_2^2 + dist_t \right) \qquad \textit{(Eq. 7)}$$

$$\text{s.t.} \quad (\mathcal{O} \cap \zeta_t^p) = \emptyset \qquad\qquad\qquad \forall t = 0, \dots, h$$

$$dist_t = \min\left\{ \left\| \zeta_t^p - \zeta_i^{ESP} \right\|_2^2 \ : \ i = 1, \dots, l \right\} \quad \forall t = 0, \dots, h$$

$$d\zeta_t^p \leq v_{max}.\, dt \qquad\qquad\qquad \forall t = 0, \dots, h$$

Where $\zeta_t^p, \zeta_i^{ESP}$ and $\zeta_g^p$ denote the patient's state at time $t$, position of available external support points and desired goal state, respectively. We solve the problem for $h$ time steps and $l$ number of external support points. $\mathcal{O}$ represents the set of obstacles and the first constraint insures a collision-free path. We add dynamics constraints such as velocity limits to guarantee that the solution is within the possible range of patient velocities. $\lambda$ is the objective weight indicating the importance of having a shorter path versus moving closer to supporting points. This is one of the patient-specific parameters in the evaluation showing the fragility of the patient.

[Place Figure 5 approximately here]

After obtaining the patient trajectory from the optimization, we find the associated type of activity for each point on the trajectory based on the defined sitting zones around the initial and target objects (Figure 5 (c)). If the patient is in the sitting zone of the initial object, the type of activity is assigned as "Sit-to-Stand", and if s/he is in the sitting zone of the target object, the type of activity is assigned as "Stand-to-Sit". Otherwise it is defined as "Walking" activity. Finally, we evaluate each point on the trajectory based on the risk calculated from the room baseline evaluation and modify it by multiplying with the trajectory factors (Figure 5 (d)).

*Turning Angle*

Turning has a high impact on risk of fall for frail patients (Cloutier, Yang, Pati, & Valipoor, 2016). We use the angular velocity to calculate patient turning angle. Using the literature on fall risk (Pati et al., 2018), we define three ranges for the amount of turning: (1) no turning, (2) turning less than 45 degrees, (3) turning more than 45 degrees. We assign 0%, 20% and 40% increase in the risk of fall for each category, respectively.

*Type of Activity*

In this initial model, we  consider "Sit-to-Stand", "Stand-to-Sit" and "Walking" activities which are well studied in the literature (Pati et al., 2018; Cloutier, Yang, Pati, & Valipoor, 2016). Although the number of falls that are reported for each activity do not necessarily show an exact ratio of their effects on risk of fall, we use those reports for initial assumptions in this model. We assign a 5% increase in risk for "Sit-to-Stand", 10%  for "Stand-to-Sit" and 20%  for "Walking".

**Overall Evaluation**

With the baseline and motion evaluations, if there is no trajectory going through a cell, we use the baseline evaluation for that cell. Otherwise, we find all the points from the generated trajectories that fall into that cell and average them as the final evaluation for that cell. This results in a risk distribution across the entire room, which is the final room evaluation. We report the results for four design examples representing several typical configurations of (1) outboard-footwall, (2) inboard-footwall, (3) inboard-headwall, and (4) nested (Phiri, 2016).

## Results

For each room example, we report two main results: (1) the baseline evaluation, and (2) the final fall risk evaluations which combines the baseline evaluation with motion evaluation. We also present evaluation along some examples of the generated trajectories.

Figures 6-9 present results for each room configuration. We show the effect of static factors in the first row, the baseline evaluation in the second row, a few samples of generated trajectories with their evaluation in the third row, and the final evaluation which is a combination of static and dynamic factors in the last row of the figure. For all the rooms, we separate day and night by having ambient light in both the main section and the bathroom for day, and only having a night light in the bathroom for night. We can see the effect of lighting using these two scenarios.

These figures show the importance of the objects support. For example, the bed and toilet have a significant effect on the overall fall risk. In the model, we assume objects (I.e., furniture and fixtures) are the same among different rooms and we are only showing the differences between various configurations, not specific brands or models of the same object. Results show that using handrails near the toilet has reduced the fall risk in the bathroom although the toilet itself remains a risky area. Also, the bed is used as a supporting object, so risk of fall is decreased in the immediate vicinity of the bed. Although reports indicate many falls occur near the bedside, we believe that this may be related to other factors such as medications or orthostatic hypotension. The effects of door operation and flooring show most change in risk occurs at the transition from the patient room to the bathroom, suggesting that more support is needed in that area. Finally, comparing the baseline and final evaluations, we see the effect of patient motion. Since we have assumed higher risk during walking, we see more risk in the path from the bed to the toilet. A shorter distance means less unsupported walking distance, which leads to lower risk of fall.

[Place Figure 6, 7, 8 and 9 approximately here]

**Discussion**

In this study, we propose a computational model that examines environmental factors of a hospital room. Results from the proposed model were presented for various room configurations showing the capability of our model to provide more objective and detailed evaluations of fall risk attributes. Logically, the results are often consistent with the literature, as the model is derived from available data, for example, a reduced risk at a toilet with grab bars on each side as compared to a single grab bar on the wall. The value of the proposed model is that stakeholders can visualize the areas of risk and evaluate decisions more holistically in the context of other choices.

As an example, teams may assume that a bathroom on the headwall is "safer," While only one headwall toilet room configuration was modeled in this phase (figure 8), the presence of a proposed wall-mounted charting station between the bed and the bathroom creates an obstruction and may diminish the assumed benefit of ease of access. Additionally, the team might more easily visualize that an obstructed path along the headwall may introduce more risk than the same distance with an unobstructed path to a bathroom in a footwall configuration. While we can't currently draw such explicit conclusions from the model, we see the opportunities for continued development that would allow such evidence-based decision-making.

Those actively involved in healthcare design understand there is rarely a single 'right' answer - design decisions need to be made in the context of competing priorities. A computational model does not provide "the" answer, but it does allow for a more informed discussion. A Design Advisory Committee of seven healthcare design experts has been formed to provide feedback on the initial model. Representatives include architects and regulators who represent design consultants, owners, and regulatory bodies. While the computational model is in early stage of development, the premise of the tool and presentation of results have received positive initial feedback. This aspect of the study is beyond the scope of this paper and will be reported separately. In the remaining, we list the limitations and possible improvements in model development:

**Limitations in the Current Model Components**

**The most important issue and limitation in our approach stems from the significant gaps in available empirical studies for both identifying contributing extrinsic factors and providing structured conclusions on the real effects of these factors on fall risk.** Most of the literature on patient falls in hospitals uses the available data from hospitals. In many cases, these data are not complete and lack important information such as time and duration of activities, exact location, or environmental configuration of the room when the patient fell. Therefore, there is not enough evidence to draw reliable quantifiable effects of the built environment factors on fall risk. Here, we briefly summarize limitations in our developed model due to the gaps in the literature:

For the support level of objects, there are many studies where a specific type of object has been studies. The results from those types of studies can only be applied for conditions that are nearly the same as the original study. However, to develop models that can evaluate various conditions, we need to focus on the ability to generalize. In other words, to understand the support level of objects we need more studies on the effect of height, movability and graspability of objects regardless of their specific type.

For the effect of flooring, we need comparisons for various floor types currently used in acute care settings. While many studies compare carpet to vinyl, due to infection prevention guidance (CDC, 2019), carpet is less frequently used in newer US patient rooms (Chari et al., 2016). In addition, more evidence is needed for flooring transitions and the associated fall risk. Other features such as thresholds between surfaces or slope of the floor (such as in bathroom) also affect the fall risk. Although designers of many newer hospital rooms are well aware of trip hazards associated with changes in floor thickness, other issues (such as the minimal slope to a shower drain or specific changes in coefficient of friction) remain less well understood. This type of more fine-grained room plan information could be added to future models. With this level of detail, it is feasible that the same tool could be used for assessing the risks in patients' homes or other facilities, which may be designed with different priorities in mind.

Studies report type of activity at the time of fall, however, none of them report the total number of times each activity was performed, making it difficult to conclude that these percentages are actually associated with the level of fall risk (Pati et al., 2018; Cloutier et al., 2016).

For the purposes of modeling door operation effect, especially the patient operation of sliding doors, further studies are needed to more reliably and accurately assign risk coefficients.

**Opportunities for Model Improvement**

We have used the simplest function that can describe the high-level effects of each component. As more studies are conducted, we can substitute these simplified functions with more detail, without changing the overall structure of the model. We can also add the intrinsic effects to customize our evaluation for a certain population, or even individual.

In obtaining the final evaluation, for each grid cell, we use the mean value from the generated trajectories. However, better metrics could be used to represent the distribution without losing important information. For example, if there are two points from trajectories in a single grid cell, one with a high fall risk and one with a low fall risk, the average is moderate. However, a grid containing a high fall risk point is important, and simply taking the average loses that information.

Finally, better patient motion prediction can improve our motion evaluation and include more realistic and broader ranges of activities. We need real data from hospitals that includes patient motion in order to use machine learning techniques for motion prediction. As ongoing work, we are conducting a series of surveys and subject studies to better inform our model.

## Conclusion

The main contributions of this paper are, (i) providing a computational model that defines and visually portrays the effect of extrinsic factors on risk of fall based on current knowledge, (ii) identifying gaps in scientific knowledge in this area, and (iii) proposing future steps to address these gaps. We believe our model represents a considerable step towards providing guidance for healthcare decision makers to understand the effects of environmental interventions on fall risk, with the goal of optimal safe patient mobility in the hospital room environment.

*Table 1*

Coefficients used in the proposed model.

| Factor | Criteria | Amount of effect | Type of effect |
|---|---|---|---|
| Floor | Resilient/compliant surface (Carpet, based on data) | 0% | No change |
| | Hard surface (Tile, based on data) | 5% | Increase |
| | Transition from resilient to hard surface | 5% | Increase |
| | Transition from hard to resilient surface | 5% | Increase |
| Light | $l < 100$ lux | 7% | Increase |
| | $100$ lux $< l < 500$ lux | 3% | Increase |
| | $l > 500$ lux | 0% | No change |
| Support | $d < 0.8$ m | 20% | Decrease |
| | $0.8$ m $< d < 1.5$ m | $(44 - (30 \times d))$ % | Decrease |
| | $d > 1.5$ m | 0% | No change |
| Door | Pull/Push − Narrow | 20% | Increase |
| | Pull/Push − Wide | 10% | Increase |
| | Slide - Narrow | 7% | Increase |
| | Slide - Wide | 4% | Increase |

*Table 2*

Predefined scenarios and their frequencies used to generate trajectories for motion evaluation.

| Scenario | Frequency |
|---|---|
| Bed to patient chair/ Patient chair to bed | 4 |
| Bed to toilet/ Toilet to sink/ Sink to bed | 18 |
| Bed to main entrance door/ Main entrance door to bed | 6 |
| Bed to sofa/ Sofa to bed | 2 |
| Patient chair to toilet/ Toilet to Patient chair | 6 |
| Total | 36 |

Fall Risk Assessment

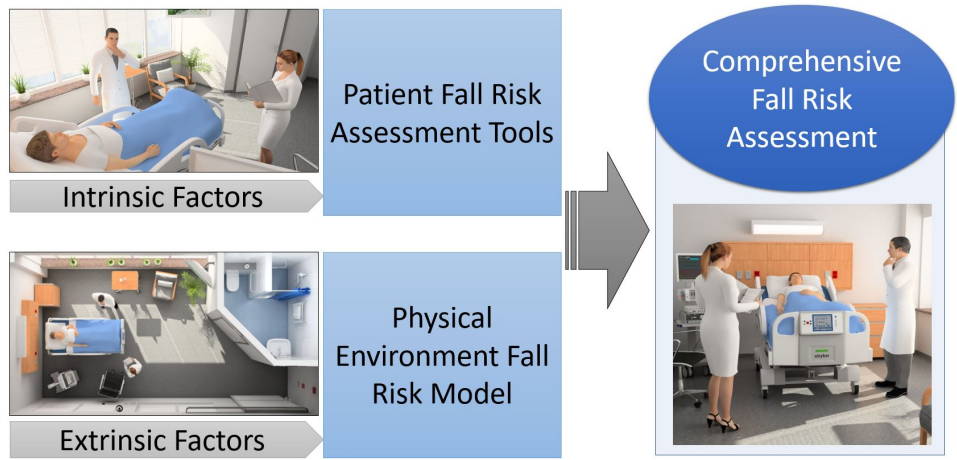

*Figure 1.* There are two aspects in fall risk assessment. In this paper we focus on physical environment and extrinsic factors contributing in risk of fall.

Fall Risk Model Flowchart

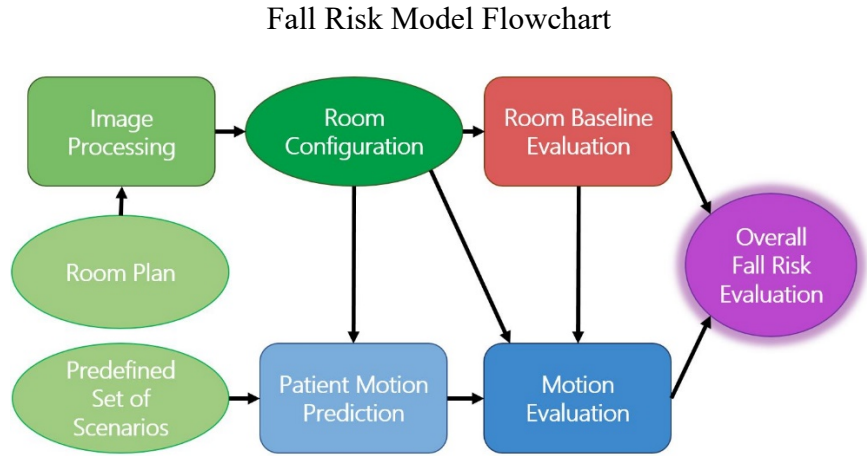

*Figure 2.* Complete fall risk evaluation model flowchart. For a given room layout image, we obtain the room configuration using image processing and calculate the room baseline evaluation – the 'static" component of risk. Based on the existing objects, we define a set of scenarios and use trajectory optimization to predict patient motion - a "dynamic" component of risk. We evaluate the risk of the predicted patient motion and combine it with the room baseline evaluation to find the overall fall risk.

Static Factors

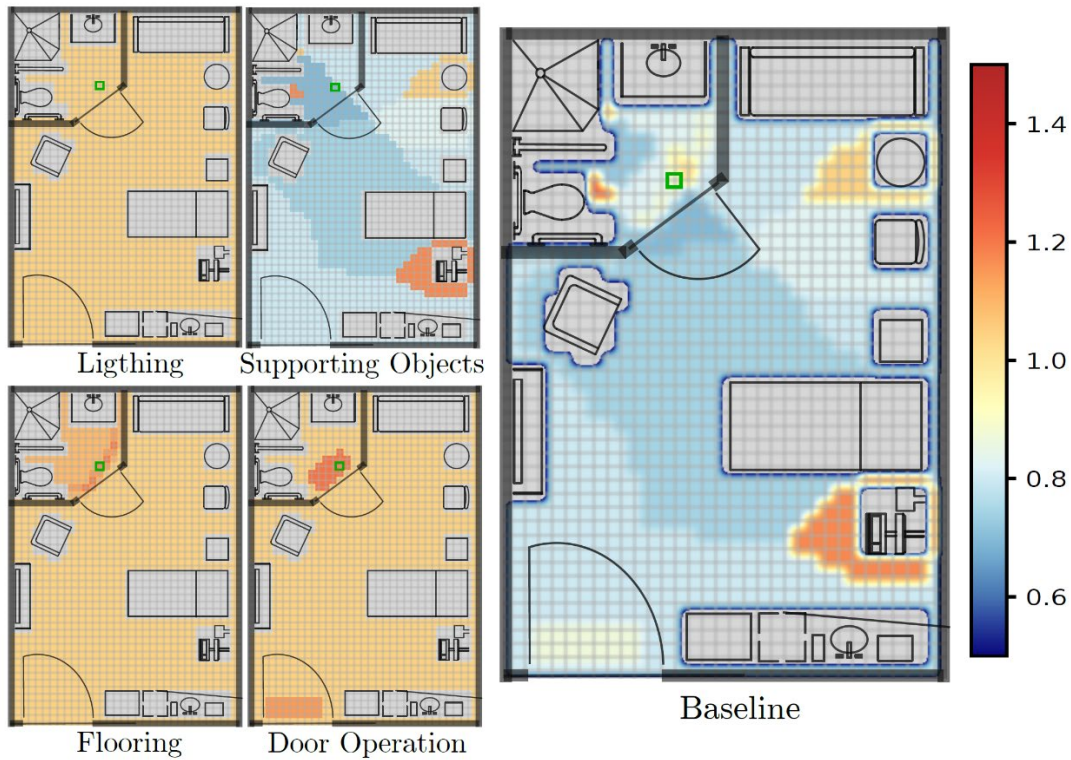

Figure 3. Static factors include floor type, lighting, door operation and the effect of supporting objects. Based on these factors, we find the room baseline fall risk evaluation which represents risk level of the entire room and is calculated as the product of all static factors. These heat maps show the percentage of increase or decrease in the fall risk. Grids in blue are considered in supportive range. Grids in yellow are considered neutral without additional support or risk. Red grids are the ones with additional risk of fall based on the extrinsic factors. Example calculation for the bolded green grid: for lighting there is a light source of 2000 lumens in the corner of the toilet which produces about 900 lux light at the example location. Since this is more than 500lux, the lighting factor is assigned as 1, meaning that it does not increase or decrease the risk of fall. For the floor surface effect, we use Eq. 2 resulting in $1 + (0.05 + 2 \times 0.05) = 1.15$. For the supporting objects, the closest supporting point is the toilet wall with distance 0.5m. The support level of walls is assigned as 1.1 since there is no handrail on that side of the wall. Since the distance is less than 0.8m, we use the first case of Eq. 6 to calculate the supporting effect which is $0.8/1.1 = 0.72$. Finally, the example grid is in the bathroom door area which is a narrow door giving it a 20% increase in risk of fall. Multiplying all these factors together gives us the baseline evaluation of $1 \times 1.15 \times 0.72 \times 1.2 = 0.99$.

Distance Function

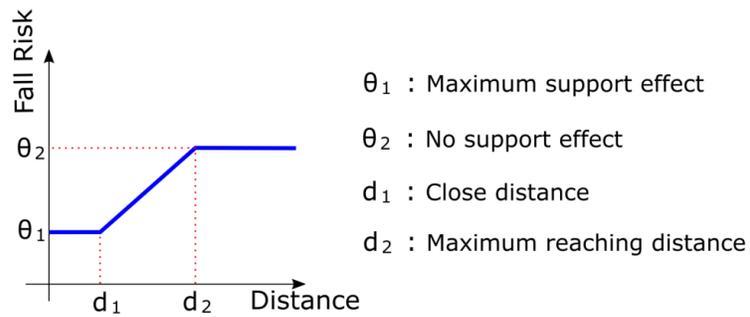

*Figure 4.* Distance function for supporting objects. The effect of a supporting object is maximum when it is close enough to the patient to grasp it without any postural change. As the distance increases, the patient has to perform some level of flexing or reaching to grasp which decreases the supporting effect of that object until it is out of reach of the patient and has no supporting effect at all.

Activity Factors

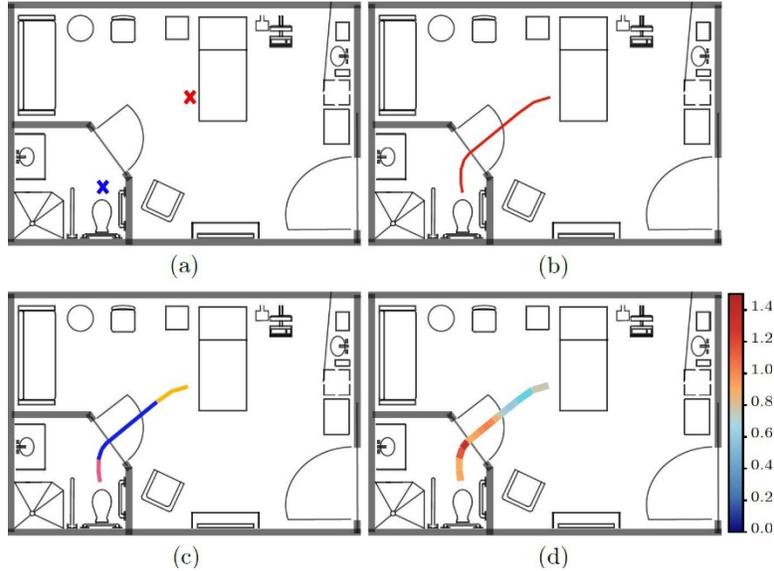

*Figure 5.* Steps in motion evaluation process for a given scenario (for example, in this figure, the scenario is "bed-to-toilet". (a) To generate each trajectory, we sample two points, one of them close to the initial location ("bed") as the start state and the other one close to the target location ("toilet") as the end state. (b) We use an optimization-based method to find a trajectory based on current state and the desired state. (c) We find the associated type of activity for each point on the trajectory based on the defined sitting zones around the initial and target objects. Here, we use green, blue and pink to present "Sit-to-Stand", "Walking" and "Stand-to-Sit" activities, respectively. (d) Finally, we evaluate each point on the trajectory based on the baseline risk evaluation and modify it by multiplying with the trajectory factors.

Evaluation of the Outboard-Footwall design

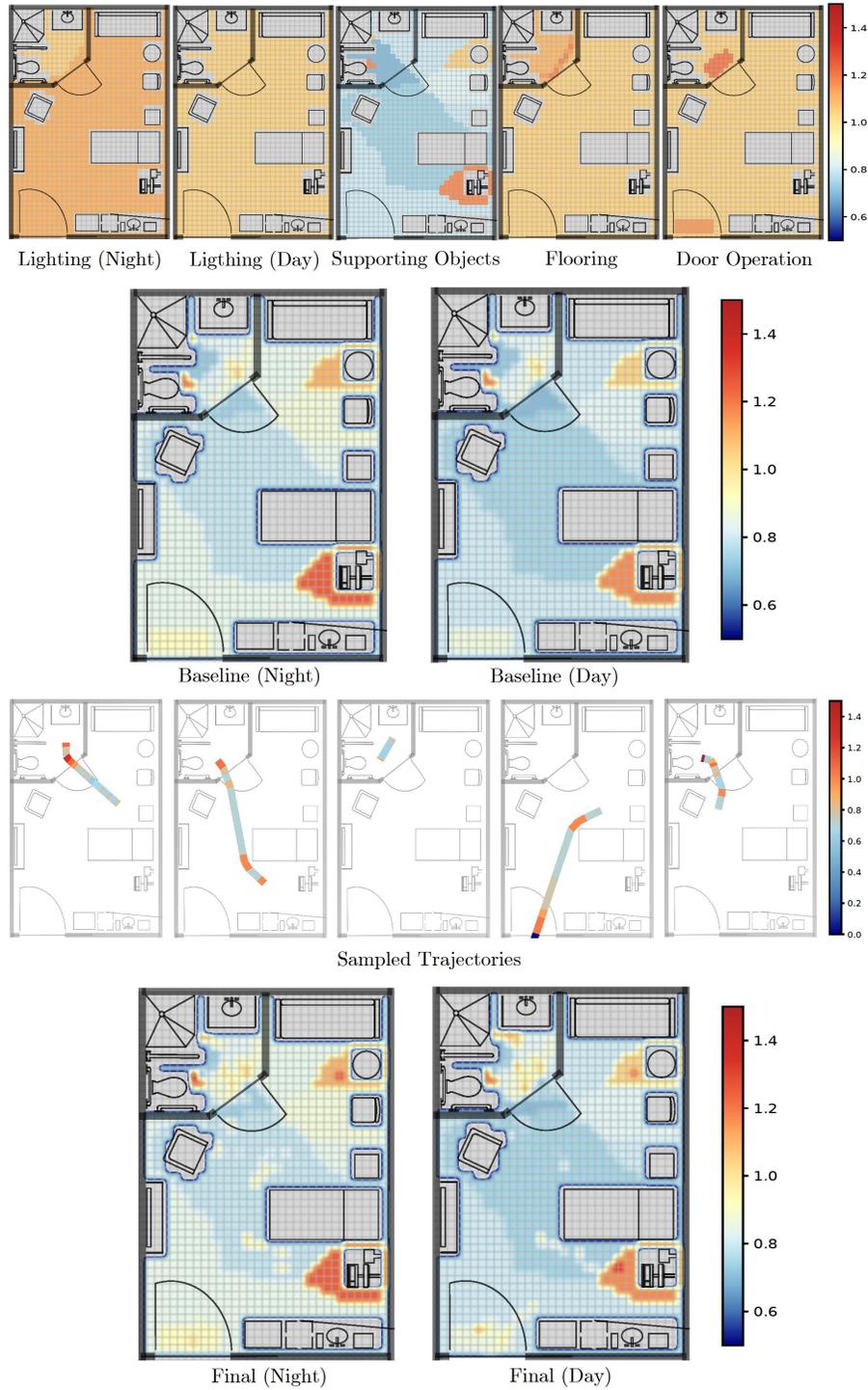

*Figure 6.* outboard-footwall room design. The first row shows the effect of each static factor separately. The second row presents the combined effect of those factors for both day and night lighting. In the third row, a few examples of generated trajectories along with their evaluations are provided. Finally, in the last row, we see the final evaluation combining the baseline evaluation and the motion evaluation.

Evaluation of the inboard-footwall design

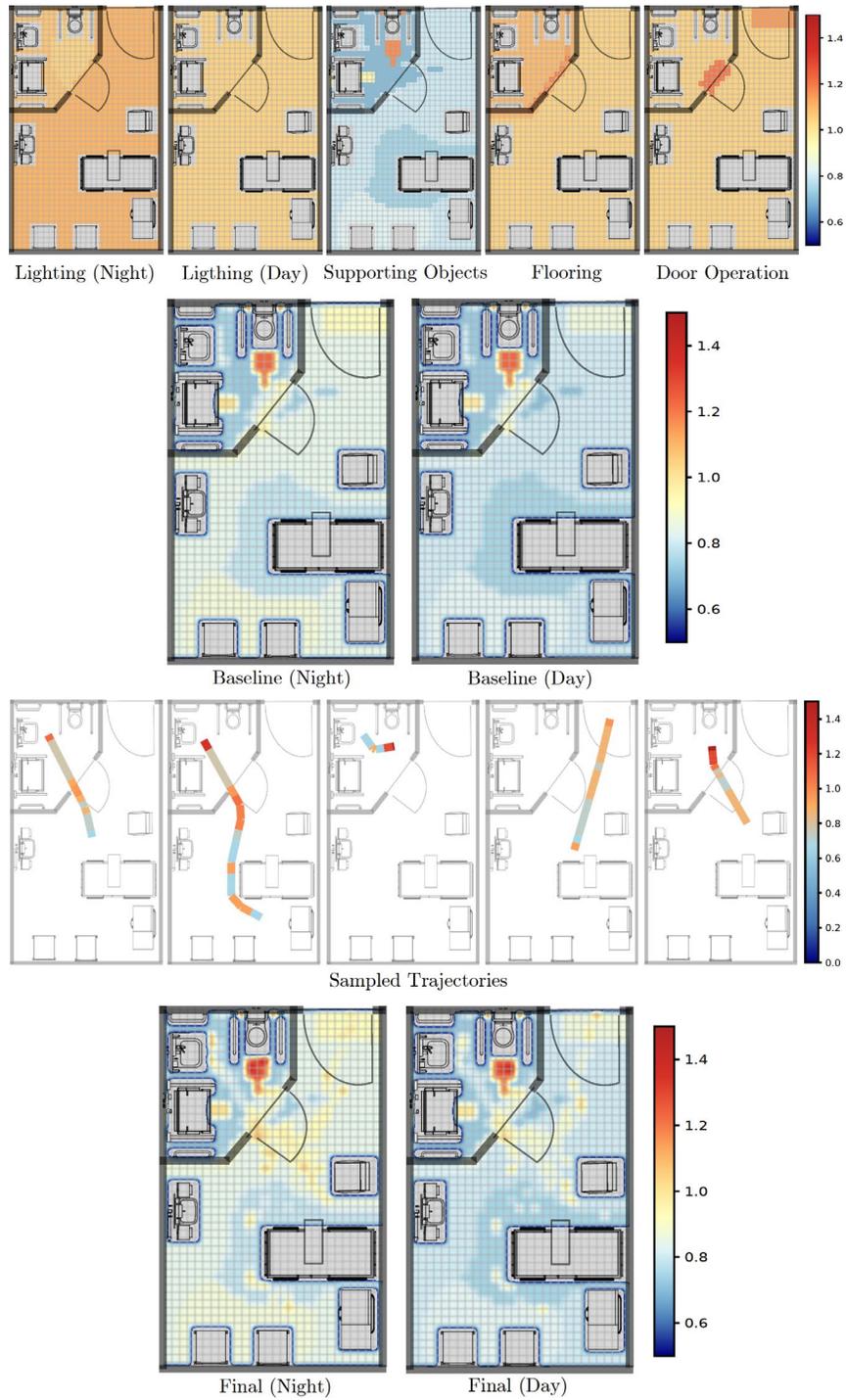

*Figure 7.* Results for the inboard-footwall design. See Figure 6 for additional description.

Evaluation of the inboard-headwall design

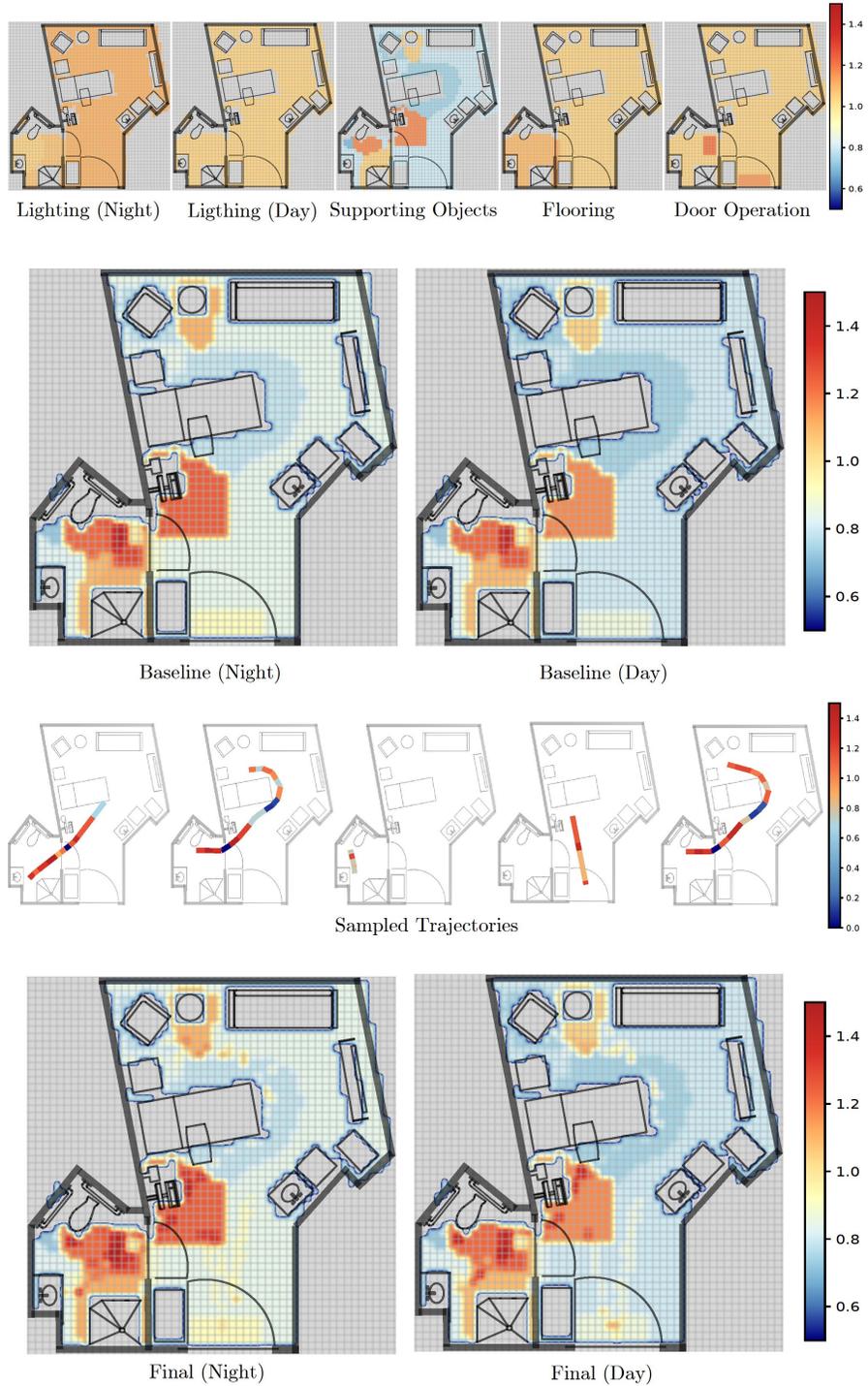

*Figure 8.* Results for the inboard-headwall design. See Figure 6 for additional description.

Evaluation of the nested design

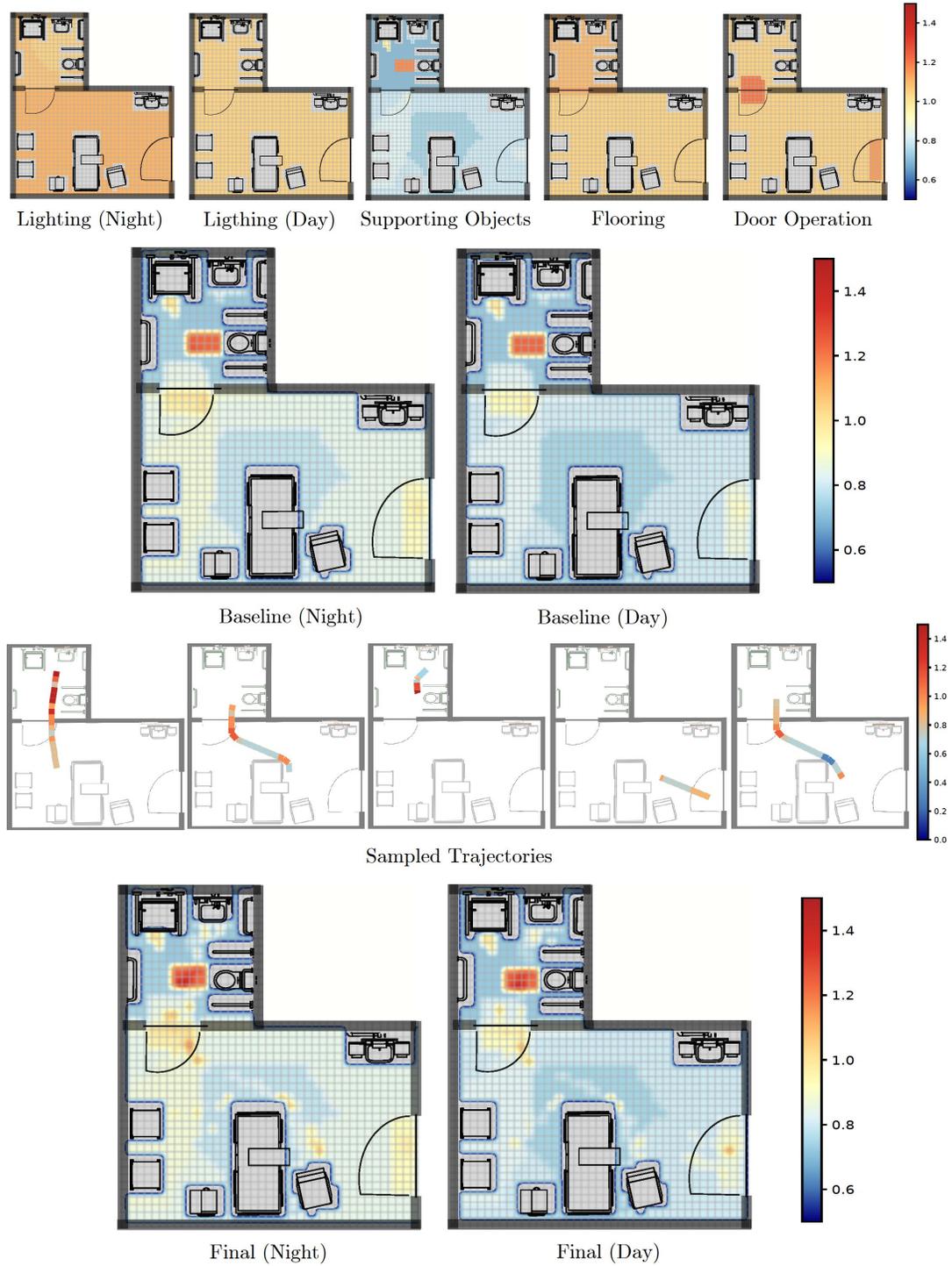

*Figure 9.* Results for the nested room design. See Figure 6 for additional description.